\documentclass[aps,pre,reprint,showpacs,nofootinbib]{revtex4-1}

\usepackage{graphicx}

\usepackage{bm}

\usepackage{amsmath,amsfonts,amssymb,amsthm}
\usepackage{latexsym}
\newcommand{\E}{\mathbb{E}}

\begin{document}

\title{Phase transition in a log-normal Markov functional model}
\author{Dan Pirjol}
\email{pirjol@mac.com}
\affiliation{Markit, 620 8$^{\rm th}$ Avenue, New York, NY 10018}

\begin{abstract} 
\noindent
We derive the exact solution of a one-dimensional Markov functional model with log-normally
distributed interest rates and constant volatility in discrete time. The model is shown to have 
two distinct limiting states, corresponding to small and asymptotically large volatilities, 
respectively. These volatility regimes are separated by a phase transition at some critical 
value of the volatility, at which certain expectation values display non-analytical behavior
as a function of volatility. We investigate the conditions under which this phase transition 
occurs, and show that it is related to the position of the zeros of an appropriately defined 
generating function in the complex plane, in analogy with the Lee-Yang theory of the phase 
transitions in condensed matter physics.
\end{abstract}
\maketitle

\section{Introduction}

An important class of interest rate models, which includes many of the models currently used in practice, 
is the class of Markov-functional models \cite{HKP,1,BH}. The advantage of these models is that the 
value of discount bonds can be expressed as a functional of a low-dimensional Markov process. 
The specification of this functional dependence allows one to model the distribution of the 
forward rates with a prescribed probability distribution. This implies that these models can 
be calibrated exactly to a set of market instruments, such as caplets or swaptions.

We consider a one-dimensional Markov-functional model with log-normally distributed forward 
Libors in discrete time. We show that the model can be solved exactly for a constant 
(time-independent) Libor volatility, and exact results can be found for the dependence of all
discount bonds on the Markovian driver. 

The exact solution of the model is used to study its behavior as a function of volatility. 
We show that the model has two distinct regimes, in the low and large volatility limits, 
respectively. These regimes are separated by a sharp transition, occurring at some critical 
value of the volatility. 
We investigate the nature of this transition, and discuss the conditions under which it occurs.
This volatility transition is similar to a first order phase transition in condensed matter 
physics, and is described by an analog of the Lee-Yang theory of phase transitions \cite{ES,LY}.

\section{The model definition}

We consider a Markov functional model with discrete time evolution. The tenor structure is
a finite set of dates 
\begin{eqnarray}
0 = t_0 < t_1 < \cdots < t_n
\end{eqnarray}
representing maturities equally spaced, e.g. by 3 or 6 months apart.

The fundamental dynamical quantities are the zero coupon bonds $P_{i,j}\equiv P_{t_i,t_j}$. 
They are functions of a one-dimensional Markov process $x(t)$, which will be assumed to 
be a simple Brownian motion with the usual properties $\E[x(t)] = 0\,, \E[x^2(t)]= t$.
The model is defined by the probability distribution of the forward Libor rates $L_{i}(t_i) = \frac{1}{\tau_i} (P^{-1}_{i,i+1} - 1)$ for the $(t_i,t_{i+1})$ period, with 
$\tau_i \equiv t_{i+1} - t_i$. We will work throughout in the $t_n-$forward measure, with numeraire
the discount bond $P_{t,t_n}$. Specifically, in this measure the Libor rates
$L_{i}$ will be assumed to be log-normally distributed 
\begin{eqnarray}\label{Li}
L_i = \tilde L_i \exp\Big( \psi x_i - \frac12 \psi^2 t_i \Big)
\end{eqnarray}
For notational simplicity we denote the value of the Markov driver at time $t_i$ as $x_i \equiv x(t_i)$. We assume that the Libor
volatility $\psi$ is a constant, although a more general formulation of the model is possible,
wherein $\psi$ has term structure. For the purpose of illustrating the phenomenon considered here,
it will be sufficient to consider a constant Libor volatility $\psi$.

We denoted in Eq.~(\ref{Li}) with $\tilde L_{i}$ the convexity-adjusted Libors; 
they are the expectation values of the Libor rates  in the measure considered, and their 
determination is part of the solution of the model to be discussed below.

The continuous time limit of this model follows the short rate process 
\begin{eqnarray}\label{conttime}
\frac{dr(t)}{r(t)} = \psi dx(t) + (\frac{d}{dt}\log \tilde r(t)) dt
\end{eqnarray}
where we introduced the convexity-adjusted forward short
rate $\tilde r(t)$ as the continuous time analog of $\tilde L_{i}$. 
The model Eq.~(\ref{conttime})
describes a simple log-normally distributed short rate model without mean reversion \cite{Dothan,BM}.
As discussed in Ref.~\cite{HKP}, mean reversion can be introduced by an appropriate choice of
the time-dependence of $\psi(t)$. Models with log-normally distributed rates in discrete time have
been considered in \cite{GZ,MSS}.

Denoting the numeraire-rebased zero coupon bond prices as
\begin{eqnarray}
\hat P_{i,j} = \frac{P_{i,j}}{P_{i,n}}
\end{eqnarray}
we note that the martingale condition for $\hat P_{i,j}$ can be expressed as
\begin{eqnarray}
\hat P_{i,j} = \E[\frac{1}{P_{j,n}}| {\cal F}_i]
\end{eqnarray}

Two particular cases of this relation are
\begin{eqnarray}\label{rec0}
&& \hat P_{i,i+1} = \E[\hat P_{i+1,i+2}(1 + \tilde L_{i+1} \tau_{i+1} f_{i+1}(x))|{\cal F}_i] \\
&& \hat P_{0,i} = \E[\hat P_{i,i+1} (1 + \tilde L_{i} \tau_i f_i(x))]
\end{eqnarray}
where we defined $f_i(x) = \exp(\psi x_i - \frac12 \psi^2 t_i)$. 

These relations can be solved recursively for $\hat P_{i,i+1}$ and $\tilde L_{i}$, 
starting with the initial conditions
\begin{eqnarray}
\tilde L_{n-1} \tau_{n-1} = \hat P_{0,n-1} - 1\,,\qquad
\hat P_{n-1,n} = 1
\end{eqnarray}
and proceeding backwards in time.
In the next section we present a method for solving these 
recursion relations in analytical form.

\begin{figure}
\begin{center}
\includegraphics[height=50mm]{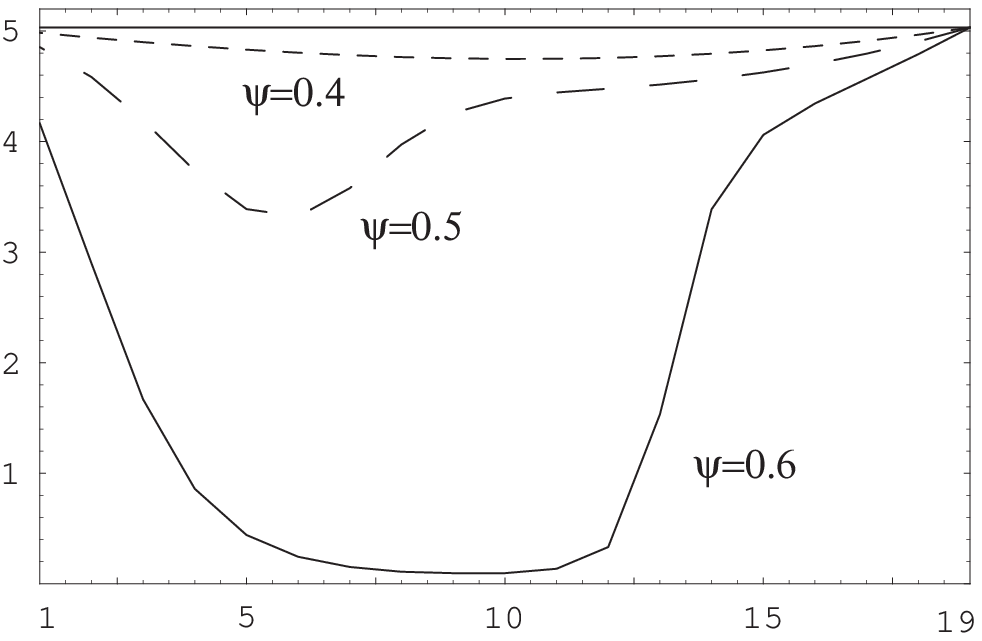}
\includegraphics[height=50mm]{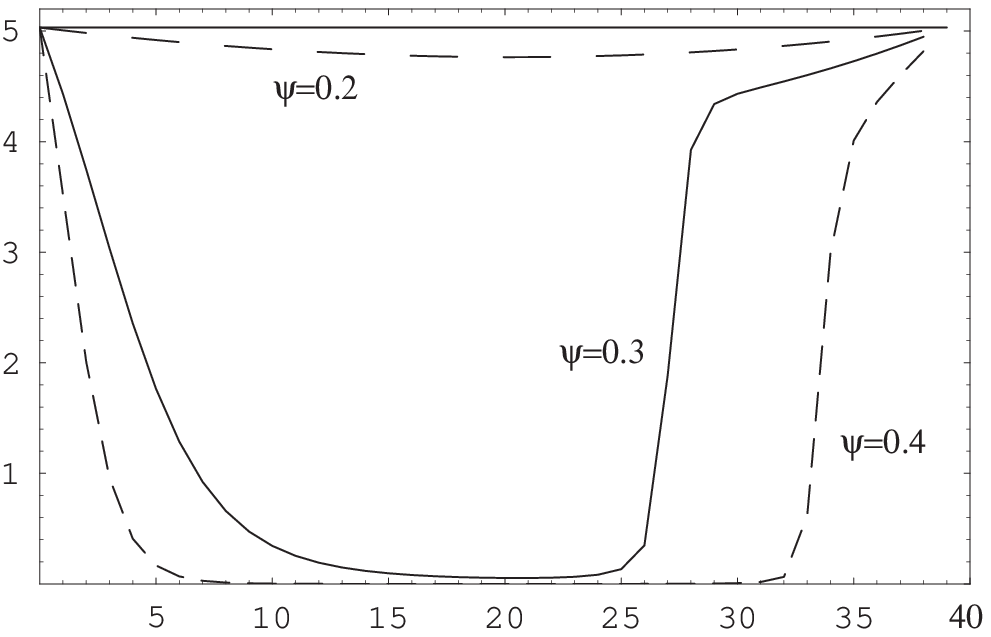}
\end{center}
\caption{The solution of the model for the convexity-adjusted Libors $\tilde L_{i}$ for a simulation with
$n=20$ time steps (above), $n=40$ (below), $\tau = 0.25$ and constant forward short rate $r_0=5\%$, for several values of the volatility $\psi$.}
\label{fig:Ltilde}
\end{figure}

\section{Analytical solution}

In practice the expectation values in Eqs.~(\ref{rec0}) are computed by numerical integration over
the functional dependence of $\hat P_{i,i+1}$ on the Markovian driver $x_i$, which is defined by appropriate interpolation. 
However, for time-independent volatility $\psi$, it can be shown that the model can be solved analytically. In this case  the solution of the recursion equation for the one-step rebased zero coupon bonds $\hat P_{i,i+1}$ has the general form
\begin{eqnarray}
\hat P_{i,i+1}(x_i) = \sum_{j=0}^{n-i-1} c_j^{(i)} e^{j\psi x_i - \frac12 (j\psi)^2 t_i}
\end{eqnarray}
with $c_j^{(i)}$ a set of constant coefficients.
The convexity-adjusted Libors are given by
\begin{eqnarray}\label{LtildeNi}
&& \tilde L_{i} = \frac{\hat P_{0,i} - \hat P_{0,i+1}}{N_i\tau _i}\,,\\
&&  N_i \equiv \E[\hat P_{i,i+1} f_i(x_i)] 
= \sum_{j=0}^{n-i-1} c_j^{(i)} e^{j \psi^2 t_i} \nonumber\,.
\end{eqnarray}

The matrix of coefficients $c_j^{(i)}$ has a triangular form (e.g. for $n=5$)
\begin{eqnarray}
\hat c = \left(
\begin{array}{ccccc}
c_0^{(n-1)} & 0 & 0 & 0 & 0 \\
c_0^{(n-2)} & c_1^{(n-2)} & 0 & 0 & 0 \\
c_0^{(n-3)} & c_1^{(n-3)} & c_2^{(n-3)} & 0 & 0 \\
\cdots & \cdots & \cdots & \cdots & \cdots \\
c_0^{(1)} & c_1^{(1)} & c_2^{(1)} & c_3^{(1)} & 0 \\
c_0^{(0)} & c_1^{(0)} & c_2^{(0)} & c_3^{(0)} & c_4^{(0)}  \\
\end{array}
\right)
\end{eqnarray}

The coefficients $c_j^{(i)}$ satisfy the recursion relation
\begin{eqnarray}\label{recursionci}
c_j^{(i)} = c_j^{(i+1)} + \tilde L_{i+1} \tau_{i+1}c_{j-1}^{(i+1)} e^{(j-1) \psi^2 t_{i+1}}
\end{eqnarray}
which must be solved simultaneously with Eq.~(\ref{LtildeNi}) for $\tilde L_i$. The initial
condition is $c_0^{(n-1)}=1, \tilde L_{n-1}\tau_{n-1} = \hat P_{0,n-1}-1$. The recursion relation (\ref{recursionci})
can be solved backwards in time, for all $i \leq n-1$, finding all coefficients in the
matrix $\hat c$ starting from the upper left corner and going downwards.

Once the coefficients $c_j^{(i)}$ and the convexity-adjusted Libors have been determined,
all zero coupon bonds can be found as
\begin{eqnarray}\label{13}
P_{i,j}(x_i) = \frac{\hat P_{i,j}(x_i)}{\hat P_{i,i+1}(x_i) [1+\tilde L_{i} \tau_i f_i(x_i)]}
\end{eqnarray}
where
\begin{eqnarray}
\hat P_{i,j} (x_i) &=& \mathbb{E}[\frac{1}{P_{j,n}}|{\cal F}_i] = 
\mathbb{E}[\hat P_{j,j+1} (1 + \tilde L_{j} \tau_j e^{\psi x_j - \frac12 \psi^2 t_j}) | {\cal F}_i]\nonumber  \\
&=& \sum_{k=0}^{n-j-1} c_k^{(j)} e^{k\psi x_i - \frac12 (k\psi)^2 t_i}\\
&+&
\tilde L_{j}\tau_j \sum_{k=0}^{n-j-1} c_k^{(j)} e^{(k+1)\psi x_i - \frac12 (k^2+1) \psi^2 t_i + 
k\psi^2(t_j - t_i)}\,.\nonumber
\nonumber
\end{eqnarray}
This completes the exact solution of the model.

The coefficients $c_j^{(i)}$ satisfy certain general relations and sum rules, valid for arbitrary volatility.
The first two coefficients $c_{0,1}^{(i)}$ can be given in closed form
\begin{eqnarray} 
&& c_0^{(i)} = 1 \mbox{  for } i = n-1, n-2, \cdots , 0 \\
&& c_1^{(i)} = \sum_{j=i+1}^{n-1} \tilde L_{j}\tau_j\,.
\end{eqnarray}
A sum rule for the coefficients which will be useful in the following is
\begin{eqnarray}
\label{sr}
&& \sum_{j=0}^{n-i-1} c_j^{(i)} = \hat P_{0,i+1}\,.
\end{eqnarray}

We illustrate the solution of the recursion in Figure~\ref{fig:Ltilde}, where we show results for the convexity adjusted Libors $\tilde L_i$ for several values
of the Libor volatility $\psi$. The two numerical examples considered assume a flat forward short rate $r_0=5\%$ and a time discretization with quarterly time steps $\tau=0.25$. The terminal bond maturity is $t_n=5$ yr, and $t_n=10$ yr, corresponding to 
$n=20, 40$ steps, respectively. 

From Fig.~\ref{fig:Ltilde} one observes that the convexity adjustment $L_{i}^{\rm fwd} - \tilde L_{i}$ 
is always positive, and increases with the volatility $\psi$. It is largest in the middle of the 
simulation interval, and with increasing volatility it becomes larger in a wider region expanding 
towards the beginning and the end of the simulation interval.  This general behavior is expected on 
general grounds for the convexity adjusted rate $\tilde L_i$ provided that $L_i$ is positively 
correlated with the
Libor associated with the payment delay $L_{t_{i+1},t_n}$. The convexity adjusted rate appears to 
vanish in the middle of the simulation interval, for sufficiently large volatilities. The rate of 
vanishing has a sudden increase for volatilities larger than a certain value. This phenomenon
is the main subject of this paper, and will be explained and quantified in Section~\ref{sec:transition} 
below.

\subsection{Scaling}

The model is uniquely defined by the parameters $\{ t_i\}, \{P_{0,i}\}, \psi$. 
The zero coupon bonds $P_{0,i}$ can be equivalently written in terms of the
zero rates $r_i$ as $P_{0,i} = \exp(-r_i t_i)$.

The model is invariant under a simultaneous rescaling of these parameters given by
\begin{eqnarray}\label{scaling}
&& t_i \to \lambda t_i\\
&& r_i \to \lambda^{-1} r_i  \nonumber \\
&& \psi \to \lambda^{-1/2} \psi \,.\nonumber
\end{eqnarray}
Under this transformation, the coefficients $c_j^{(i)}$ and the expectation values $N_i$ are invariant, 
while the convexity adjusted Libors scale as $\tilde L_i \to \lambda^{-1} \tilde L_i$.
Due to this scaling invariance, the number of the relevant parameters of the 
model is reduced from three to two.

\subsection{Generating function}

In this section we present an efficient method for solving
the recursion relation (\ref{recursionci}) for the coefficients $c_j^{(i)}$.
We introduce the generating function at the time horizon $t_i$
\begin{eqnarray}
f^{(i)}(x) \equiv \sum_{j=0}^{n-i-1} c_j^{(i)} x^j
\end{eqnarray}
This function satisfies a recursion relation, expressing the generating function at time
$t_i$ in terms of the generating function at the next time $t_{i+1}$
\begin{eqnarray}
f^{(i)}(x) = f^{(i+1)}(x) + \tilde L_{i+1} \tau x f^{(i+1)}(x e^{\psi^2 t_{i+1}})
\end{eqnarray}
The initial condition for the recursion is $f^{(n-1)}(x) = 1$.
The expectation value $N_i$ appearing in the expression for the convexity-adjusted Libor 
$\tilde L_{i}$ Eq.~(\ref{rec0}) is
\begin{eqnarray}\label{Nif}
N_i = f^{(i)} (e^{\psi^2 t_i} )
\end{eqnarray}

The generating function $f^{(i)}(x)$ takes known values at $x=0,1$
\begin{eqnarray}\label{fi01}
f^{(i)}(0) = 1\,,\qquad
f^{(i)}(1) = \hat P_{0,i+1}
\end{eqnarray}
where the second relation follows from the sum rule Eq.~(\ref{sr}).

In the zero volatility limit $\psi=0$, the generating function $f^{(i)}(x)$ can be found exactly
\begin{eqnarray}\label{fi0}
f^{(i)}_0(x) = \Pi_{j=i+1}^{n-1} (1 + L_{j}^{\rm fwd} \tau_j x)\,.
\end{eqnarray}
where $L^{\rm fwd}_j = \tau_j^{-1}(\hat P_{0,j}/\hat P_{0,j+1} - 1)$ is the 
forward Libor rate for the time period $(t_j, t_{j+1})$.
Expanding in powers of $x$ this gives all the coefficients $c_j^{(i)}$ in the zero volatility limit.

As the volatility increases $\psi > 0$, the generating function $f^{(i)}(x)$ also changes, 
in such a way that the constraints Eqs.~(\ref{fi01}) are still satisfied.

The recursion relation for $f^{(i)}(x)$ can be 
reformulated in such a way that it does not contain any reference to the convexity-adjusted
Libors 
\begin{eqnarray}
f^{(i)}(x) = f^{(i+1)}(x) + (\hat P_{0,i+1} - \hat P_{0,i+2}) x 
\frac{f^{(i+1)}(x e^{\psi^2 t_{i+1}})}{f^{(i+1)}(e^{\psi^2 t_{i+1}})}\nonumber \\
\end{eqnarray}

In the asymptotically large volatility limit $\psi \to \infty$, this recursion relation can be
again solved exactly, and the generating function $f^{(i)}(x)$ 
is given by the asymptotic form
\begin{eqnarray}\label{fiasym}
&& f^{(i)}_\infty(x) = 1 + (\hat P_{0,n-1} - 1)x + (\hat P_{0,n-2} - \hat P_{0,n-1}) x^2\nonumber\\
&& + \cdots + (\hat P_{0,i+1} - \hat P_{0,i+2})x^{n-i-1}\,.
\end{eqnarray}
Note that the coefficients $c_j^{(i)}$ have well-defined limiting values as $\psi\to \infty$.

This shows that the model considered has two very different limiting regimes, corresponding to i) small 
volatility, and ii) large volatility. We will denote these regimes as the phases of the model.
In each phase the generating function $f^{(i)}(x)$ has a well-defined expansion, given by
Eqs.~(\ref{fi0}) and (\ref{fiasym}), respectively.

We can use the results for the generating function in the zero and large
volatility limits in order to obtain asymptotic expressions for the 
convexity adjusted Libors $\tilde L_i$ in the small and large volatility
limits. This can be done using the relation between the expectation values 
$N_i$ defined in Eq.~(\ref{LtildeNi}), and the generating function 
$N_i = f^{(i)}(e^{\psi^2 t_i})$.

We start by considering first the asymptotics of $\tilde L_i$ in the
small volatility limit $\psi^2 t_i \ll 1$. The expansion of the generating
function $f_0^{(i)}(x)$ around $x=1$ reads
\begin{eqnarray}
f_0^{(i)}(x) &=& \hat P_{0,i+1}[1 + \sum_{j=i+1}^{n-1}
\frac{L_j^{\rm fwd} \tau_j}{1 + L_j^{\rm fwd} \tau_j} (x-1)\\
& & + O((x-1)^2)]\,.\nonumber
\end{eqnarray}
Using Eq.~(\ref{LtildeNi}) this gives the small volatility asymptotics
of the convexity-adjusted Libors $\tilde L_i$, valid up to 
corrections of $O((\psi^2 t_i)^2)$
\begin{eqnarray}\label{CvxAdjLow}
&& \tilde L_i \simeq L_i^{\rm fwd}(1 - \sum_{j=i+1}^{n-1}
\frac{L_j^{\rm fwd} \tau_j}{1 + L_j^{\rm fwd} \tau_j} (e^{\psi^2 t_i}-1))\\
&& \hspace{3cm} \psi^2 t_i \ll 1\nonumber
\end{eqnarray}
(Note that the use of the zero volatility limit of the generating
function $f_0^{(i)}(x)$ was sufficient in order to derive this result.
This is due to the exact condition $f^{(i)}(1)= \hat P_{0,i+1}$, which 
implies that the Taylor expansion of $f^{(i)}(x,\psi^2 t_i)$ in powers
of $x-1$ and $\psi^2 t_i$ contains only a linear term in $x-1$ but not in
$\psi^2 t_i$.)

The low volatility approximation Eq.~(\ref{CvxAdjLow}) has the familiar form of the 
convexity adjustment for a log-normally distributed rate. The growth of the convexity
adjustment with the volatility has the familiar exponential form, 
proportional to $\exp(\psi^2 t_i) - 1$. Assuming a flat forward Libor curve,
we have $L_i^{\rm fwd} =  L^{\rm fwd} \equiv \frac{1}{\tau} (e^{r_0\tau} - 1)$. 
Then all terms in the sum over $j$ are equal, and the result (\ref{CvxAdjLow})
simplifies as
\begin{eqnarray}
&& \tilde L_i \simeq L_i^{\rm fwd}(1 - (n-i-1)
\frac{L^{\rm fwd} \tau_j}{1 + L^{\rm fwd} \tau_j} (e^{\psi^2 t_i}-1))
\end{eqnarray}
This convexity adjustment is largest in the middle of the simulation
interval, and vanishes near the boundaries. This agrees qualitatively with the 
main features of the convexity adjustment observed in Figure~\ref{fig:Ltilde}. 

Consider next the large volatility asymptotics $\psi^2 t_i \gg 1$ of the
convexity adjusted Libors $\tilde L_i$. This follows from the large $x$
asymptotics of the large volatility generating function $f_\infty^{(i)}(x)$
\begin{eqnarray}\label{finfr0}
f_\infty^{(i)}(x) \to (\hat P_{0,i+1} - \hat P_{0,i+2}) x^{n-i-1}+ O(x^{n-i-2})\,.
\end{eqnarray}
Assuming again a flat forward Libor curve, this gives the large volatility
asymptotics of the convexity adjusted Libors
\begin{eqnarray}\label{CvxAdjHigh}
\tilde L_i &=&  L^{\rm fwd} \left(\frac{1 + L^{\rm fwd}\tau}{L^{\rm fwd}\tau}\right) 
e^{-(n-i-1)\psi^2 t_i} \\
& & \hspace{3cm} \psi^2 t_i \gg 1\nonumber
\end{eqnarray}
This shows that in the large volatility limit, the convexity adjusted rates $\tilde L_i$
drop off much faster with the volatility $\psi$. The decrease is still exponential, but it is
much faster due to the additional factor $n-i-1$ (equal to the number of time steps to maturity) 
in the exponent $\tilde L_i \sim \exp(-(n-i-1)\psi^2 t_i)$. The vanishing of the
convexity adjusted Libors is faster in the middle of the simulation interval,
just like in the small volatility case.

As an aside, we note that all dynamical quantities of the model can be expressed 
formally in terms of the generating 
function $f^{(i)}(x)$. For example, the rebased bond prices are given by
\begin{eqnarray}
&&\hat P_{i,i+1}(x) = \exp(-\frac12 t_i \partial_x^2) f^{(i)}(e^{\psi x})\\
&& \hat P_{i,j}(x) = \exp(-\frac12 t_i \partial_x^2) f^{(j)}(e^{\psi x}) \\
&&\qquad + \tilde L_j \tau_j e^{\psi x - \frac12 \psi^2 t_i}
f^{(j)}(e^{\psi x + \psi^2 (t_j-t_i)})\,,\nonumber
\end{eqnarray}
from which all zero coupon bond prices can be obtained using Eqs.~(\ref{13}).

\section{Libor volatility transition}\label{sec:transition}

As the volatility increases from zero to a large value, the coefficients $c_j^{(i)}$ interpolate between
the two limiting values, corresponding to low and asymptotically large volatilities, respectively.
Equivalently, the generating function $f^{(i)}(x)$ changes between the two limiting expressions
$f_0^{(i)}(x)$ and $f_\infty^{(i)}(x)$, in such a way that the two constraints Eq.~(\ref{fi01}) are still satisfied.

\begin{figure}
\begin{center}
\includegraphics[height=52mm]{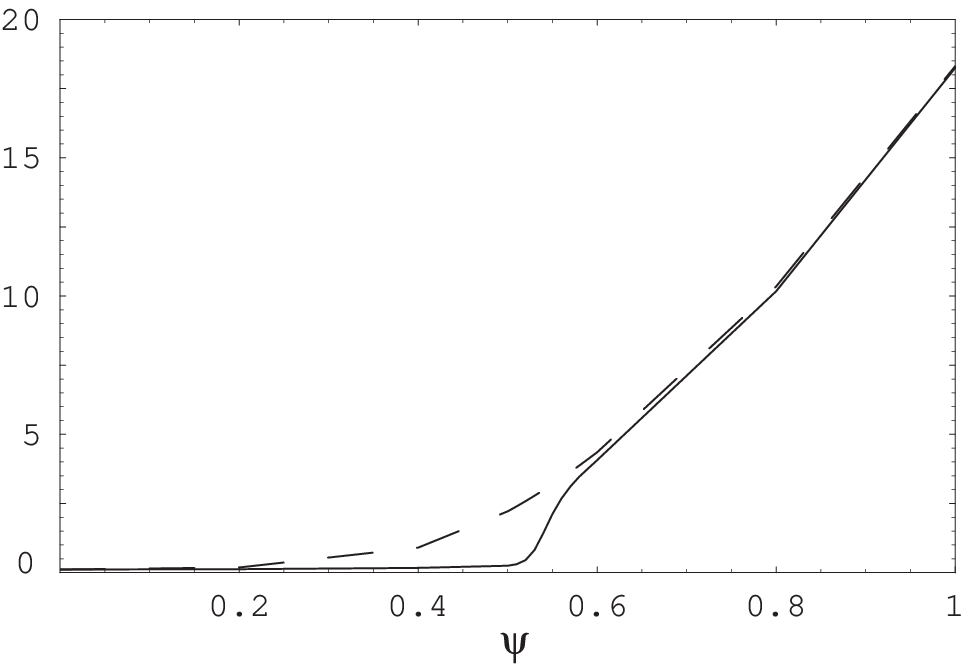}
\includegraphics[height=53mm]{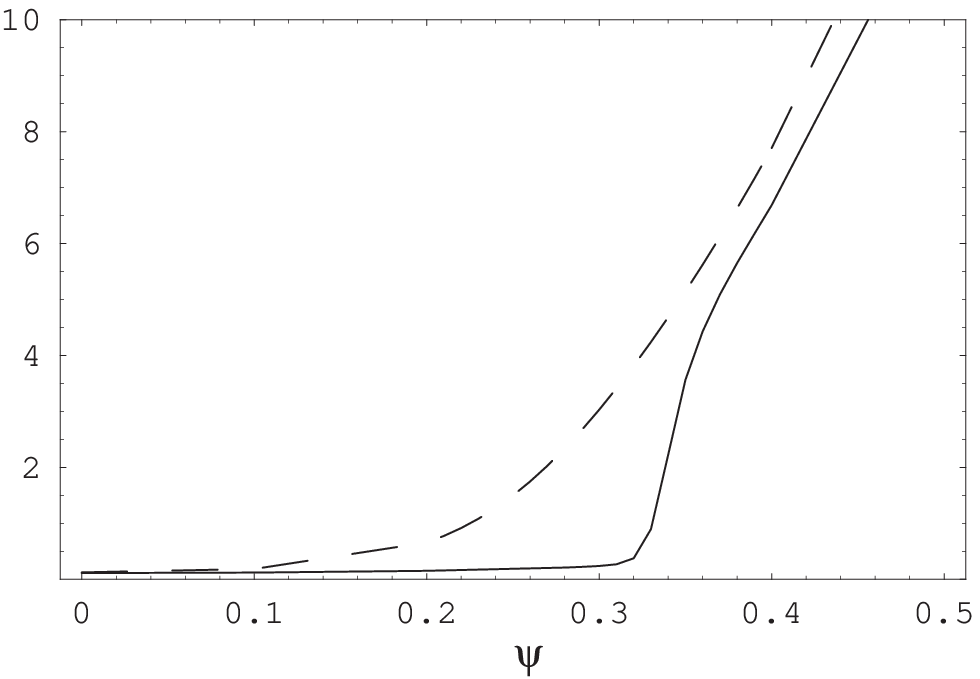}
\end{center}
\caption{The log of $N_i$ as a function of the volatility $\psi$ (solid line), and 
its large volatility  approximation as $N_i = f_\infty^{(i)}(\exp(\psi^2 t_i))$ (dashed line).
Above: $n=20, i=10$. Below: $n=40,i=30$. Both cases correspond to $\tau = 0.25, r_0=5\%$.}
\label{fig:logN10}
\end{figure}

As mentioned above, the convexity adjusted Libors $\tilde L_i$
appear to become vanishingly small at some value of the volatility, 
see Fig.~\ref{fig:Ltilde}. This phenomenon occurs first in the middle of the simulation time interval, 
and then it gradually extends also towards the boundaries. This is related to the expectation values 
$N_i$ which become very large as the volatility increases. To investigate this in more detail, 
we show in Fig.~\ref{fig:logN10} the plots of $\log N_i$ as function of the volatility $\psi$ for 
two simulations.

We observe that the change is not gradual, but happens at a sharply defined value of the 
volatility, which will be called the critical volatility $\psi_{\rm cr}$. The transition becomes more
sharp as $n-i-1$ increases.
The critical volatility 
demarcates two regions of very different qualitative  behaviour, in which the model has distinct
limiting expressions for the functional dependence of discount bonds on the Markov driver.

The convexity-adjusted Libors $\tilde L_i$ become very small for volatilities above the critical value 
$\psi > \psi_{\rm cr}$. In practice,
they can become so small that they are below machine precision. This phenomenon thus
imposes a limit to the practical applicability of the model, and it is important to understand the
conditions under which it occurs.

In the following we investigate in some detail the nature of this transition,
and formulate a criterion for finding the critical volatility $\psi_{\rm cr}$ at each time 
horizon $t_i$.
We will show that the singular behaviour of the expectation values $N_i$ is related
to the distribution of the zeros of the generating function $f^{(i)}(z)$ in the
complex plane. 

\begin{figure}
\begin{center}
\includegraphics[height=70mm]{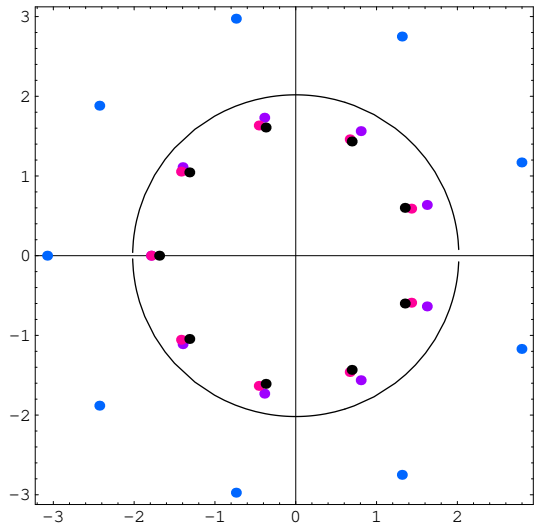}
\end{center}
\caption{The zeros of the generating function $f^{(i)}(x)$ in the complex plane for several values of the 
volatility $\psi=0.5,0.55,0.6$ around the critical value $\psi_{\rm cr}=0.53$ (blue, violet, red). 
The black dots correspond to the large volatility asymptotic generating function $f_\infty^{(i)}(x)$, 
which is obtained for $\psi=\infty$. The parameters are the same as in the
upper plot of Fig.~\ref{fig:logN10}, and the time slice $t_i$ considered is $i=10$ . The circle shown has radius $\exp(\psi_{\rm cr}^2 t_i)$.}
\label{fig:roots}
\end{figure}

The generating function $f^{(i)}(z)$ is a polynomial of $z$ with positive coefficients, and thus 
does not have any zeros on the positive real axis $z>0$. However, it is well known
that the position of the zeros in the complex plane can influence the behaviour of the
function along the real axis.
The generating function $f^{(i)}(z)$ has $n-i-1$ zeros. At zero volatility, all zeros are at large
real negative values $z_k = - 1/(L_k^{\rm fwd}\tau_k)$, but they migrate in the complex plane
as pairs of complex conjugate values as the  volatility increases, and surround the origin,
see Fig.~\ref{fig:roots}. 
As the volatility increases to very large values, the zeros reach fixed positions, given by the
zeros of the asymptotic generating function $f_\infty^{(i)}(x)$.

Furthermore, as the polynomial order $n-i-1$ increases, the number of the zeros increases
and they close in on the
positive real axis, pinching it at some point $z_* > 1$. We will show that the function $f^{(i)}(z)$ is
continuous at $z_*$, but its derivative has a jump. Thus the generating function has a cusp 
at this point. Recalling that $N_i$ is related to the generating function
as $N_i = f^{(i)}(e^{\psi^2 t_i})$, see  Eq.~(\ref{Nif}), it follows that $N_i$ has a singular
behaviour at the volatility $\psi_{\rm cr}$, given by the equation 
\begin{eqnarray}\label{volcr}
\exp(\psi_{\rm cr}^2 t_i) = z_*\,.
\end{eqnarray} 
This equation determines the critical volatility $\psi_{\rm cr}$ at the time slice $t_i$. Geometrically, this 
has the following meaning: the critical volatility at the time horizon $t_i$ is given by that
value of $\psi$ for which the zeros of the generating function $f^{(i)}(z)$ enter the
circle of radius $\exp(\psi^2 t_i)$. As $\psi$ increases, the zeros move closer to the origin,
while the circle of radius $\exp(\psi^2 t_i)$ expands, such that at some intermediate value $\psi_{\rm cr}$, 
the zeros will cross the expanding circle. This picture is illustrated in Figure \ref{fig:roots} on the 
example of the transition shown in the upper plot of Figure \ref{fig:logN10}.

The transition between the two volatility phases is 
similar to a phase transition in the Lee-Yang formalism \cite{LY}, where critical points are
associated with the values of the thermodynamic parameter (fugacity) at which the
zeros of the grand canonical partition function pinch the real axis. For a general introduction to phase
transitions see \cite{ES}.
As shown in \cite{LY}, the partition function is continuous at the critical point, but its derivative
has a jump, which is proportional to the density of zeros around this point. These results hold in
the thermodynamical limit of an infinite volume; the analog of the thermodynamical limit in 
our case is $n-i-1 \to \infty$, where $n-i-1$ is the number of time steps from maturity to the time
slice considered $t_i$.

Such a behaviour is precisely what is observed in Fig.~\ref{fig:logN10}, where one can see that $\log N_i$ is continuous everywhere, but its derivative has a jump at $\psi_{\rm cr}$.
The analog of the partition function in our case is the generating function $f^{(i)}(z)$, and the 
jump occurs at $z_* = \exp(\psi^2_{\rm cr}t_i)$
In the following we quantify this statement, and compute an explicit result for the discontinuity
of the derivative of $f^{(i)}(z)$ at $z_*$.

The generating function can be written explicitly in terms of its zeros $z_k$ as 
\begin{eqnarray}
f^{(i)}(z) = \Pi_{k=1}^{n-i-1} (1 - z/z_k)
\end{eqnarray}

As the polynomial order $n_f=n-i-1$ increases, the roots arrange themselves on a closed curve
around the origin, which can be parameterized in polar coordinates as $z(\theta) = \rho(\theta) e^{i\theta}$. The roots appear in complex conjugate pairs, which implies that the curve
describing the zeros is symmetric under reflection on the real axis $\rho(\theta) = \rho(-\theta)$.

The logarithm of the generating function $f^{(i)}(z)$ can be expressed in the limit $n_f \to \infty$ as an integral
\begin{eqnarray}\label{logfi}
\log f^{(i)}(z)\! =\! \int_0^{\pi} d\theta g(\theta) \log(\frac{z^2}{\rho^2(\theta)} - 2\cos\theta \frac{z}{\rho(\theta)} + 1)
\end{eqnarray}
where $g(\theta)$ is the density of roots at polar angle $\theta$. It is normalized as
\begin{eqnarray}
\int_0^\pi d\theta g(\theta) = \frac12 n_f
\end{eqnarray}

The derivative of $\log f^{(i)}(z)$ is 
\begin{eqnarray}
\frac{d}{dz} \log f^{(i)}(z)\! =\! 2 \int_0^{\pi} d\theta g(\theta) \frac{z - \rho(\theta)\cos\theta}{z^2 - 2\rho(\theta)\cos\theta z + \rho^2(\theta)}
\end{eqnarray}

This is discontinuous across the boundary at $z = \rho(0)$ with a jump given by the density of zeros at
this point
\begin{eqnarray}\label{derivjump}
&& \frac{d}{dz} \log f^{(i)}(z)|_{z = \rho(0) + \epsilon} - 
\frac{d}{dz} \log f^{(i)}(z)|_{z = \rho(0) - \epsilon}\nonumber\\
&& \qquad\qquad = 2\pi g(0) \frac{1}{\rho(0)}\,.
\end{eqnarray}
These results are completely analogous to the expressions derived in \cite{LY} for the jump of the
derivative of the grand canonical function at a critical point.

A similar phenomenon occurs for any expectation value of the form similar to $N_i$, with $\phi$
a real number
\begin{eqnarray}
\mathbb{E}[\hat P_{i,i+1} e^{\phi x - \frac12 \phi^2 t_i}] = f^{(i)} (e^{\psi\phi t_i})
\end{eqnarray}
The expectation value can be expressed in terms of the generating function $f^{(i)}(x)$ as shown.
The critical volatility corresponding to this expectation value is found in analogy to
Eq.~(\ref{volcr}) and is given by $\exp(\psi\phi t_i) = z_*$.

\begin{figure}
\begin{center}
\includegraphics[height=55mm]{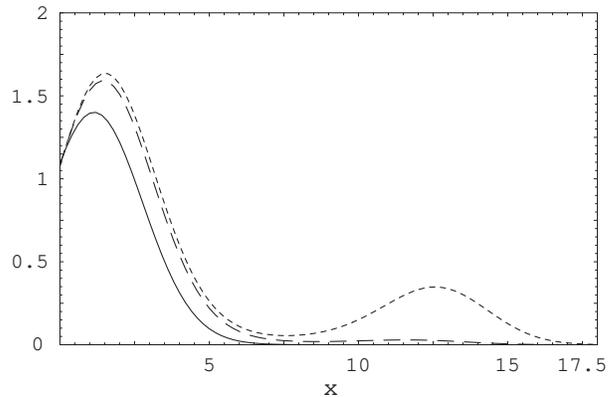}
\end{center}
\caption{The integrand in the expression (\ref{integrand}) for the expectation value $N_i$ at time
$t_i=2.5$ for 
several values of the volatility $\psi$: i) 0.4 (solid line), 0.5 (dashed line), 0.52 (dotted line).
The parameters of the simulation are $n=20,i=10$, $r_0=5\%, \tau=0.25$. Note the second maximum
which appears for volatilities close to the critical volatility $\psi_{\rm cr}$.}
\label{fig:integrand}
\end{figure}

We close this section with a comment about the practical observability of the phase
transition in usual implementations of the Markov functional model. The expectation
value $N_i$ defined in (\ref{LtildeNi}) is written explicitly as
\begin{eqnarray}\label{integrand}
N_i &=& 
\mathbb{E}[\hat P_{i,i+1} f_i(x) ] \\
&=& 
\int_{-\infty}^\infty \frac{dx_i}{\sqrt{2\pi t_i}} e^{-\frac{x_i^2}{2t_i}} \hat P_{i,i+1}(x_i) e^{\psi x_i - \frac12 \psi^2 t_i}\nonumber
\end{eqnarray}
This integral is usually assumed to be dominated by contributions from small values of $x_i$
around the origin $|x_i| \leq \kappa \sqrt{t_i}$, where $\kappa \sim 3-5$. The integral is then evaluated
numerically either on a grid, or using Monte Carlo simulations. However, for volatilities $\psi$
above the critical value, the integrand develops a second local maximum at large values of the
Markov driver as seen in Figure~\ref{fig:integrand}, which will dominate the integral above the
critical volatility. In the example of Figure~\ref{fig:integrand}, the secondary maximum appears at 
$x\sim 12$, which is almost 10 standard deviations away from zero. 
Thus the integral (\ref{integrand}) will receive significant
contributions from a region in $x$ which is sampled very inefficiently in Monte Carlo or grid
methods. This implies that the model will not be simulated correctly in the large volatility phase,
and the phase transition will be unobservable under these simulation methods.

\section{Example: constant forward short rate}

We illustrate the general results discussed in the previous section on the example of a forward yield curve with constant short rate $r_0$. The initial yield curve is $P_{0,i} = \exp(- r_0 t_i)$, and the
numeraire rebased discount bonds are $\hat P_{0,i} = \exp(r_0 (n-i)\tau)$.

In Figure~\ref{fig:roots} we show the zeros $z_k$ of the generating function $f^{(i)}(z)$ at the
time slice $i=10$, for a simulation with $r_0=5\%$, total simulation time
$t_n=5$ yr, with time step $\tau = 0.25$ and $n=20$ time steps. The colored dots show the exact 
zeros at three values of the volatility $\psi = 0.5, 0.55, 0.6$ around the critical volatility $\psi_{\rm cr}=0.53$, and the black dots are the zeros of the asymptotic generating function $f_\infty^{(i)}(z)$. 

We note that the asymptotic generating function $f_\infty^{(i)}(z) $ gives a reasonably good approximation for the position of the zeros in the large volatility phase. For this reason we discuss in some detail the position of the zeros of the asymptotic generating function $f_\infty^{(i)}(z) $
for which an analytical treatment is possible in the limit of a constant forward short rate.

The asymptotic generating function  for a constant forward short rate $r_0$ is
\begin{eqnarray}
f_\infty^{(i)}(x) = 1 + [1 - e^{-r_0\tau}] \sum_{j=1}^{n-i-1} (e^{r_0\tau} x)^j
\end{eqnarray}
Its zeros are $x_k = e^{-r_0 \tau} z_k$, where $z_k$ are the zeros of the polynomial $p_{n_f}(z)$ of
degree $n_f=n-i-1$
\begin{eqnarray}\label{finfz}
p_{n_f}(z) \equiv \frac{1}{1-e^{-r_0\tau}} +  z + z^2 + \cdots + z^{n_f} 
\end{eqnarray}
Under usual market conditions $1/(1-\exp(-r_0 \tau)) > 1$, and by the Enestr\"om-Kakeya theorem \cite{BP}, all zeros of this polynomial lie outside the open unit disk $|z_k| > 1$.

The structure of the zeros of $p_{n_f}(z)$ can be 
studied by noting that this polynomial is the truncated Taylor series of the function
\begin{eqnarray}
F(z) = \frac{1}{e^{r_0\tau} - 1} + \frac{1}{1-z}
\end{eqnarray}
The function $F(z)$ has an exact zero at $z_0 = e^{r_0\tau}$, and a pole at $z=1$, which 
means that the convergence region is the circle $|z| < 1$. 

The theory of the zeros of a truncated Taylor
expansion is a well studied subject in approximation theory. The main result is the
Jentzsch-Szeg\"o theorem \cite{J, Sz}, according to which the zeros of the truncated Taylor series of a 
function $F(z)$ either converge to the zeros of $F(z)$ as $n_f\to \infty$, provided that they are inside its
convergence region, or they accumulate on the boundary of the convergence region. 
For our case the latter situation applies, such that the zeros of $p_{n_f}(z)$ will accumulate uniformly on the $|z|=1$ circle as the polynomial order $n_f$ increases.

This implies that the zeros of $f_\infty^{(i)}(x)$ close in on the positive real axis at $x_* = \rho \equiv e^{-r_0\tau}$
as the polynomial order $n_f = n-i-1$ becomes large. This phenomenon is visible already at moderate values of $n_f \sim O(10)$, as seen from Fig.~\ref{fig:roots}.

We can compute the properties of the generating function at the critical point $x_*$ by
applying the general results discussed above. Taking
$\rho(\theta) = \rho \equiv e^{-r_0 \tau}$ and $g(\theta) =
n_f/(2\pi)$, the integral in Eq.~(\ref{logfi}) can be performed exactly with the result
\begin{eqnarray}
\log f_\infty^{(i)}(x) = 
\left\{
\begin{array}{ll}
n_f \log \rho\,, & x \leq \rho \\
n_f \log x\,, & x \geq \rho \\
\end{array}
\right.
\end{eqnarray}

The jump of the derivative across the critical point $x_* = \rho$ is
\begin{eqnarray}
\frac{d}{dx} \log f_\infty^{(i)}(x = \rho + \epsilon) - 
\frac{d}{dx} \log f_\infty^{(i)}(x = \rho - \epsilon) \!=\!
\frac{n_f}{\rho}
\end{eqnarray}
In the large volatility phase the derivative of $\log f^{(i)}_\infty(x)$ is 
very large, and is of the same order of magnitude 
as expected from the asymptotic form Eq.~(\ref{finfr0}).

Finally, we consider the case of practical interest of finite polynomial order $n_f=n-i-1$.
It was observed in \cite{CM} that for the truncated Taylor series $a_0 + a_1 z + \cdots + a_n z^n$ of a function $F(z)$,
a good approximation for the moduli of the zeros $|z_k|$ is obtained
by neglecting all but the first and last terms (by considering the simpler polynomial
 $\tilde p(z) \equiv a_0 + a_n z^{n}$), provided that the function $F(z)$ does not have zeros within the convergence region. In our case of the polynomial $p_{n_f}(z)$, 
the corresponding polynomial $\tilde p_{n_f}(z) = 1/(1-e^{-r_0 \tau}) + z^{n_f}$
has zeros
\begin{eqnarray}
z_k &=& (\frac{1}{1- e^{-r_0 \tau}})^{1/(n-i-1)} e^{\pi i \frac{2k-1}{n-i-1}}\,,\\
& & \qquad \qquad k = 1, 2, \cdots , n-i-1\nonumber
\end{eqnarray}

The critical volatility can be computed using equation Eq.~(\ref{volcr}) and is given by
\begin{eqnarray}\label{volcr2}
&& e^{r_0\tau + \psi_{\rm cr}^2 t_i} =  (\frac{1}{1- e^{-r_0 \tau}})^{1/(n-i-1)} \,,\\
&&\psi_{\rm cr}^2 \simeq \frac{1}{i(n-i-1)\tau} \log(\frac{1}{r_0\tau} )\,.\nonumber
\end{eqnarray}
The minimum value of $\psi_{\rm cr}$ is reached at $i=[n/2]$, in the middle of the simulation
interval, where $i(n-i-1)$ is maximal. This agrees with the shape of the $\tilde L_i$ curves in
Fig.~\ref{fig:Ltilde}, where the critical volatility is first reached in the middle of the interval.
Thus the practical applicability range of the model is restricted to volatilities smaller than the
minimum critical volatility 
\begin{eqnarray}
\psi^2 < (\psi_{\rm cr}^2)_{\rm min} =  \frac{1}{[n/2]^2\tau} \log(\frac{1}{r_0\tau} )\,.
\end{eqnarray}
This expression satisfies the general scaling properties of the model Eqs.~(\ref{scaling}).

\begin{table*}
\begin{center}
\begin{tabular}{|c|c|c|c|c|c|c|c|c|}
\hline
& \multicolumn{2}{|c|}{$t_n=5$} & \multicolumn{2}{|c|}{$t_n=10$} & \multicolumn{2}{|c|}{$t_n=20$} & \multicolumn{2}{|c|}{$t_n=30$} \\
\hline
$r_0$ & $\tau = 0.25$ & $\tau = 0.5$ & $\tau = 0.25$ & $\tau = 0.5$ & $\tau = 0.25$ & $\tau = 0.5$ & $\tau = 0.25$ & $\tau = 0.5$ \\
\hline
\hline
$1\%$ & 48.95\% & 65.10\% & 24.48\% & 32.55\% & 12.24\% & 16.28\% & 8.16\% & 10.85\% \\
$2\%$ & 46.04\% & 60.70\% & 23.02\% & 30.35\% & 11.51\% & 15.17\% & 7.67\% & 10.12\% \\
$3\%$ & 44.24\% & 57.96\% & 22.12\% & 28.98\% & 11.06\% & 14.49\% & 7.37\% & 9.66\% \\
$4\%$ & 42.92\% & 55.94\% & 21.46\% & 27.97\% & 10.73\% & 13.99\% & 7.15\% & 9.32\% \\
$5\%$ & 41.87\% & 54.32\% & 20.93\% & 27.16\% & 10.47\% & 13.58\% & 6.98\% & 9.05\% \\
\hline
\end{tabular}
\caption{The maximal Libor volatility $\psi$ for which the model is everywhere below the critical
volatility $\psi_{\rm cr}$, determined according to Eq.(\ref{volcr2}), for several choices of the total
tenor $t_n$, time step $\tau$ and the level of the interest rates $r_0$ .}
\end{center}
\label{table1}
\end{table*}

In Table~1 we show the values of the maximally allowed volatility for several values of the total simulation tenor $t_n = n\tau$, the time step $\tau$, and the short rate $r_0$. The maximum 
allowed volatility decreases
with the size of the time step $\tau$, with the tenor of the simulation $t_n$, and with 
the short rate $r_0$.

\section{Conclusions}

We discussed in this paper the behaviour of a Markov functional model with discrete log-normally distributed Libors, as a function of the Libor volatility. The model can be solved exactly in the $T-$forward measure in the limit of
a time-independent volatility, by a backwards
recursion relation. Analytical results can be obtained for the functional dependence of all
discount bonds on the Markov driver.

We showed that the model has two volatility regimes, corresponding to small and large volatility,
with very different qualitative behaviour. As the Libor volatility increases, there is a transition between
the two regimes, at an intermediate critical volatility. In the large volatility phase the convexity
adjusted Libors are very small, and can be below machine precision. Thus the existence of the large
volatility regime imposes a limit on the applicability of such a model, which is manifested as an 
upper bound on the allowed Libor volatility.

We formulated the conditions under which
this  phase transition occurs, and showed that it is related to the position of the zeros of an appropriately 
defined generating function in the complex plane. The transition appears for long simulation times, and
small time discretization steps. A similar phenomenon is expected to occur also for the practically
relevant but analytically more complex case of time-dependent volatility $\psi(t_i)$.
Also, the discussion of this paper is limited to the $T-$forward measure, and one expects that the details of the volatility dependence could change in a different measure, but not the existence of a phase transition. We hope to report progress along these directions in future work.
\vspace{5pt}

\appendix
\section{The zeros of the generating function $f_\infty^{(i)}(x)$}

We study here the distribution of the complex zeros of the asymptotic generating function $f_\infty^{(i)}(z)$ for a constant forward short rate $r_0$. This is related to the problem of finding
the zeros of the polynomial
\begin{eqnarray}\label{pnf}
p_{n}(z) \equiv a +  z + z^2 + \cdots + z^{n} 
\end{eqnarray}
where $a$ is a real number larger than 1. They are the same as the zeros of the 
equation $z^{n+1} + (a-1) z - a = 0$ from which $z=1$ is excluded.
Expressed in polar coordinates $z = \rho e^{i\theta}$ one finds that the zeros are on the curve
$\rho(\theta)$ which is given by the solution to the equation
\begin{eqnarray}\label{rhoeq}
\rho^{2(n+1)}(\theta) \!=\! (a-1)^2 \rho^2(\theta) - 2 a(a-1) \rho(\theta) \cos\theta + a^2
\end{eqnarray}

\begin{figure}
\begin{center}
\includegraphics[height=70mm]{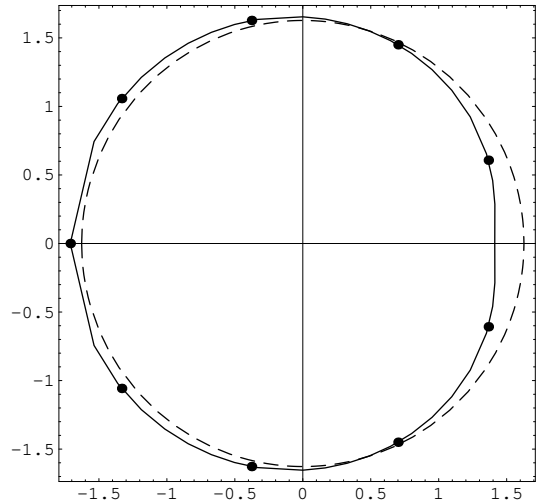}
\end{center}
\caption{The zeros of the polynomial $p_9(z)$ for $a=80$. The solid line denotes the
curve $\rho(\theta)$ given in Eq.~(\ref{rhoeq}), and the dashed line shows the approximation for the 
moduli of the zeros $|z_k| = a^{1/n}$ obtained
by retaining in $p_n(z)$ only the first and last terms. }
\label{fig:app}
\end{figure}

Figure \ref{fig:app} shows the exact roots of $p_9(z)$, along with the curve $\rho(\theta)$ (solid line),
and the circle with radius $a^{1/n}$ (dashed line), which was used in the main text. We note that the latter approximates the moduli of the zeros very well, as noted in \cite{CM}.

The curve $\rho(\theta)$ intersects the real axis at some point $\rho(0)$. The solution of the equation
(\ref{rhoeq}) at $\theta=0$ has always the solution $\rho(0)=1$, and in addition it can have another 2 solutions, or none, depending on the values of $n,a$. The solutions different from 1 can be 
found as the solutions of the simpler equation $h(\rho) = \rho^{n+1} - (a-1) \rho + a = 0$. The function $h(\rho)$ has a minimum at $\rho_* = [(a-1)/(n+1)]^{1/n}$. 
We distinguish the 3 cases, according to the value of $h(\rho_*)$
\begin{eqnarray}
1. && h(\rho_*) < 0 \to \mbox{ 2 solutions for $\rho$} \\
2. && h(\rho_*) =0 \to \rho = \rho_* \\
3. && h(\rho_*) > 0 \to \mbox{ no solutions for $\rho$} 
\end{eqnarray}
These cases are obtained for $n<[n_*], n=[n_*], n>[n_*]$ respectively, where $n_*$ is the solution
of the equation 
\begin{eqnarray}
(\frac{a}{n})^n = (\frac{a-1}{n+1})^{n+1}\,.
\end{eqnarray}
For $a=80$ one has $[n_*]=22$. A good approximation for the solutions in case 1 is obtained by
Taylor expanding $h(\rho)$ around $\rho_*$ to quadratic order. This gives
\begin{eqnarray}
\rho = \rho_* \pm \sqrt{2 \frac{n\rho_*^{n+1} - a}{n(n+1)\rho_*^{n-1}}}\,.
\end{eqnarray}
In Fig.~\ref{fig:app} we show only the largest of the 3 solutions.

The angular distribution of the roots can be obtained from the imaginary part of the equation
for $z$, which reads $\rho^n \sin[(n+1)\theta] = - (a-1) \sin\theta$. For the subset of even $n$,
this has real solutions for $\rho$ only if $\theta$ is in one of the regions
\begin{eqnarray}
\theta \in ((2k-1)\frac{\pi}{n+1}, 2k\frac{\pi}{n+1})\,,\quad k = 0, 1, \cdots n-1
\end{eqnarray}
There is one solution in each of these intervals of equal angular opening, which means that in the
large $n$ limit, the angular distribution of the roots approaches a uniform distribution 
$g(\theta) = n/(2\pi)$. In  the same limit, the curve $\rho(\theta)$ approaches the unit 
circle $\rho(\theta) = 1$, as required by the Jentzsch-Szeg\"o theorem.

\section*{Acknowledgements}

I am grateful to Dyutiman Das and Adrian Ghinculov for useful discussions on this problem, and 
to Radu Constantinescu for comments on the manuscript.


\end{document}